\documentclass[]{aa}

\usepackage{graphicx}
\usepackage{txfonts}
        
\begin{document}

\title{Evolution and colors of helium-core white dwarf stars with 
       high-metallicity progenitors}

\author{L. G. Althaus$^{1,4}$\thanks{Member of the Carrera del Investigador
        Cient\'{\i}fico y Tecnol\'ogico, CONICET, Argentina.},
        J. A. Panei$^{1,2}$, 
        A. D. Romero$^{1,2}$\thanks{Fellow of CONICET, Argentina.}, 
        R. D. Rohrmann$^{3\star}$, 
        A. H. C\'orsico$^{1,2\star}$, 
        E. Garc\'{\i}a--Berro$^{4,5}$, and
        M. M. Miller Bertolami$^{1,2\star\star}$}

\offprints{L. G. Althaus}

\institute{$^1$Facultad de Ciencias Astron\'omicas y Geof\'{\i}sicas,
               Universidad Nacional de La Plata,
               Paseo del Bosque S/N, 
               (B1900FWA)La Plata, Argentina.\\
           $^2$Instituto  de Astrof\'{\i}sica de La Plata, 
               IALP (CCT La Plata), 
               CONICET-UNLP\\
           $^3$Observatorio Astron\'omico, 
               Universidad Nacional de C\'ordoba, 
               Laprida 854 (5000), 
               C\'ordoba, Argentina \\
           $^4$Departament de F\'\i sica Aplicada, 
               Escola Polit\`ecnica Superior de Castelldefels,
               Universitat Polit\`ecnica de Catalunya,  
               Av. del Canal Ol\'\i mpic, s/n,  
               08860 Castelldefels,  
               Spain\\
           $^5$Institute for Space Studies of Catalonia,
               c/Gran Capit\`a 2--4, 
               Edif. Nexus 104,   
               08034 Barcelona,  
               Spain\\
\email{althaus@fcaglp.unlp.edu.ar} }

\date{Received; accepted}

\abstract{}
         {Motivated  by  the recent  detection  of  single and  binary
          He-core white  dwarfs in  metal-rich clusters, we  present a
          full set of evolutionary calculations and colors appropriate
          for the study of such  white dwarfs. The paper is also aimed
          at  investigating   whether  stable  hydrogen   burning  may
          constitute a main source of energy for massive He-core white
          dwarfs resulting from high-metallicity progenitors.}
         {White dwarf sequences are derived by taking into account the
          evolutionary  history of  progenitor  stars with  supersolar
          metallicities.   We  also  incorporate   a  self-consistent,
          time-dependent  treatment   of  gravitational  settling  and
          chemical  diffusion,  as well  as  of  the residual  nuclear
          burning.}
         {We  find  that the  influence  of  residual nuclear  burning
          during the late stages  of white dwarf evolution is strongly
          dependent  on the  occurrence of  chemical diffusion  at the
          base  of the  hydrogen-rich envelope.  When no  diffusion is
          considered,  residual hydrogen  burning  strongly influences
          the  advanced  stages of  white  dwarf cooling,  introducing
          evolutionary  delays  of  several  Gyr.  By  contrast,  when
          diffusion is taken into account the role of residual nuclear
          burning is strongly mitigated, and the evolution is dictated
          only  by  the  thermal  content  stored  in  the  ions.   In
          addition,  for all  of  our sequences,  we provide  accurate
          color and  magnitudes on the  basis of new and  improved non
          gray model atmospheres  which explicitly include   Ly$\alpha$   
          quasi-molecular   opacity.}
         {}


\authorrunning{L. G. Althaus et al.}

\titlerunning{Evolution   of   helium-core   white  dwarfs  with  high 
              metallicity progenitors.}

\maketitle


\section{Introduction}

White  dwarfs   constitute  the  most  common   end-point  of  stellar
evolution.  In fact, more than 95$\%$ of all stars are expected to end
their lives as white dwarfs.  As such, the present population of white
dwarfs conveys  information about  the history of  our Galaxy  and has
potential applications as reliable cosmic clocks to infer the age of a
wide variety  of stellar populations  --- see Winget \&  Kepler (2008)
for a recent review.

The mass distribution of hot  white dwarfs (Kepler et al.  2007) peaks
at about $0.59  \, M_{\sun}$ and exhibits a  significant low-mass tail
of  white dwarfs  with masses  smaller than  $0.45\,  M_{\sun}$, which
peaks  at $0.40\,  M_{\sun}$.   These low-mass  white  dwarfs are  the
result of strong mass-loss episodes  during the Red Giant Branch (RGB)
evolution  before  the onset  of  the  helium  core flash.  For  solar
metallicity stars  this is  thought to occur  only in  binary systems.
The fact that low-mass He-core white dwarfs are predominantly found in
close binary systems --- mostly  with another white dwarf or a neutron
star companion (van Kerkwijk et  al.  2005) --- supports the idea that
most of  these white  dwarfs are the  result of  mass-tranfer episodes
during binary  evolution. Here, a companion strips  the outer envelope
from a post main-sequence star before  the star reaches the tip of the
RGB. Accordingly,  and due to  the relative simplicity of  white dwarf
evolution, He-core white  dwarfs have been used to  constrain the ages
and properties of millisecond pulsars (van Kerkwijk et al. 2005).

The  interest  in low-mass  white  dwarfs  is  also motivated  by  the
discovery  of candidate  He-core  white dwarfs  in  open and  globular
clusters.  We mention the open cluster M67 (Landsman et al. 1997), the
globular cluster NGC 6397 --- where the first He-core white dwarf in a
globular  cluster was  identified (Cool  et al.   1998; Taylor  et al.
2001) --- and 47  Tuc (Edmonds  et al. 2001).   More recently,  it has
been suggested that the excess of very hot (extreme) horizontal branch
stars in $\omega$Cen  and in NGC 2808 --- which are  assumed to be the
result of strong mass-loss episodes on the RGB --- could be reflecting
the existence of a sizeable  fraction of He-core white dwarfs in these
Galactic globular  clusters (Castellani et al.  2007;  Calamida et al.
2008).  In fact, as discussed in Calamida et al. (2008), He-core white
dwarfs have been detected in  stellar systems that harbor an important
fraction  of extreme  horizontal branch  stars, thus  implying  that a
fraction of RGB stars might avoid the helium core flash (Castellani et
al. 1994).  The presence of  He-core white dwarfs in these clusters is
the result of extreme mass  loss possibly caused by stellar encounters
in the  high stellar  density environment of  globular clusters  or by
evolution in  compact binaries.  Thus,  the existence of  an important
population  of   He-core  white  dwarfs  offers   the  possibility  of
constraining globular cluster dynamics  and evolution (Moehler \& Bono
2008).

The possible  detection of {\sl  single} low-mass white dwarfs  in NGC
6791,  one of the  oldest ($\ga  8$ Gyr)  and most  metal-rich ([Fe/H]
$\approx +0.4$) open clusters in our Galaxy (Origlia et al. 2006), has
opened a vivid debate.   Kalirai et al.  (2007) have spectroscopically
identified in the upper part of the cooling sequence of NGC 6791 a few
very bright  white dwarfs whose masses are  apparently consistent with
single  massive He-cores.  The  mean  mass of  these  white dwarfs  is
$\langle  M\rangle =0.43 \pm  0.06\, M_{\sun}$.   The existence  of an
important population of massive He-core white dwarfs in this extremely
metal-rich cluster was suggested by Kalirai et al. (2007) on the basis
that  NGC 6791  contains  an important  number  of extreme  horizontal
branch stars. These He-core white dwarfs and extreme horizontal branch
populations would be, according to  Kalirai et al.  (2007), the result
of enhanced mass losses on the  RGB in this cluster, perhaps driven by
its high metal content. However, Van Loon et al.  (2008) have recently
presented evidence that  there is no indication for  the occurrence of
strong mass  loss among the  stars in NGC  6791.  Owing to  the longer
cooling  times of  He-core  white dwarfs,  a  substantial fraction  of
massive He-core white  dwarfs was invoked by Hansen  (2005) to explain
the observed bright peak in the white dwarf luminosity function of NGC
6791,  although this  has been  recently  challenged by  Bedin et  al.
(2008a)  who claim that  the peak  in the  luminosity function  can be
accounted for by an important  population of double white dwarf binary
systems in the cluster.  Finally, based on theoretical arguments about
the   initial-final   mass  relationship   for   stars  of   different
metallicities, Meng et al.  (2008) have shown that metal-rich low-mass
stars  may  become  undermassive  white dwarfs.   According  to  these
authors,  a  population  of   undermassive  white  dwarfs  ($M<0.5  \,
M_{\sun}$) are expected in metal-rich old clusters.

A proper interpretation of the observations of He-core white dwarfs in
high  metallicity   environments  like  that  of   NGC  6791  requires
evolutionary calculations.  In  particular, it requires computing full
evolutionary  sequences of  white dwarfs  with  supersolar metallicity
progenitors.  However, a detailed study of the evolutionary properties
of  He-core  white  dwarfs  resulting from  progenitors  with  extreme
metallicity does  not exist  in the literature.   The present  work is
intended to  fill this shortcoming.  Specifically, we  have computed a
full  set  of  evolutionary  calculations  for  He-core  white  dwarfs
covering a  wide range of stellar  masses. We have  taken into account
the  evolutionary  history  of   progenitor  stars  for  the  case  of
supersolar metallicities  of $Z$=0.03, 0.04,  and 0.05.  {\bf  The pre
white dwarf evolution has been simulated by removing mass from a $1 \,
M_{\sun}$  model at  the  appropriate stages  of  its evolution.}   We
present evolutionary calculations  which include a self-consistent and
time-dependent  treatment of  gravitational settling  and  of chemical
diffusion as  well as a careful  study of the role  played by residual
nuclear burning. In  this particular, one of the aims  of this work is
to investigate whether stable hydrogen burning may constitute the main
source of energy for evolved  He-core white dwarfs resulting from high
metallicity  progenitors.   In addition,  for  all  our sequences,  we
provide  accurate colors  and  magnitudes based  on  new and  improved
non-gray  model   atmospheres  which  explicitly   include  Ly$\alpha$
quasi-molecular  opacity  according   to  the  approximation  used  by
Kowalski  \& Saumon  (2006).  In  view  of the  possibility that  some
He-core white  dwarfs in  NGC 6791 could  have masses larger  than the
canonical value  of the  helium core mass  at the helium  flash (about
$0.45  \, M_{\sun}$  at the  high metallicity  of NGC  6791),  we have
extended  the mass  range of  our calculations  computing evolutionary
sequences  for  He-core  white  dwarfs  with masses  up  to  $0.55  \,
M_{\sun}$.   To   this  end,  we  have   artificially  suppressed  the
occurrence of the  He-core flash at the tip of  the RGB.  The possible
existence  of He-core  white  dwarfs above  $\sim  0.50 \,  M_{\sun}$,
possibly due to rotation, has  been recently discussed by Bedin et al.
(2008b) as an alternative to explain some observed features of the RGB
and cooling sequences in NGC 6791.  The paper is organized as follows.
In Sect.  2 we comment  on the main existing evolutionary calculations
for He-core  white dwarfs.  Sect.   3 contains details about  the main
ingredients of our evolutionary calculations and of the non-gray model
atmosphere  we  use  to  compute  the colors  and  magnitudes  of  our
evolutionary  sequences.   In  Sect.    4  we  describe  our  results,
emphasizing  the role of  nuclear burning  and element  diffusion.  In
this  section we  also discuss  the the  evolution of  massive He-core
white dwarfs. In Sect. 5 we  briefly comment on the predictions of our
model atmospheres for the white  dwarf sequences.  Finally, Sect. 6 is
devoted to discuss and summarize our results.

\section{Evolutionary calculations for He-core white dwarfs}

The study  of the evolution of  low-mass He white  dwarfs has captured
the attention of researchers  since Kippenhahn et al. (1967) suggested
that these stars could be  the natural result of substantial mass loss
from  low-mass red  giant stars  filling their  Roche lobes  in binary
systems.  Later, Webbink (1975) was  the first to show that for masses
larger than  $0.17\, M_{\sun}$, He-core white  dwarfs develop hydrogen
shell flashes.   He found that the  cooling of these  white dwarfs was
significantly affected  by the contribution  of residual proton-proton
(pp) hydrogen  burning in the envelope.  Long  cooling times resulting
from residual hydrogen burning were also reported by Castellani et al.
(1994).   These authors  considered  the evolutionary  history of  the
progenitor  stars and  clearly demonstrated  that for  massive He-core
white dwarfs  (even close to the  upper limit for the  mass of He-core
white  dwarfs), the contribution  of pp  burning can  sensitively slow
down  the cooling  up  to  log$({L/L_{\sun}}) \sim  -4$  and that  for
smaller  masses the cooling  is progressively  affected by  strong CNO
flashes, the occurrence  of which depends on the  metal content of the
parent star.

Althaus  \&  Benvenuto  (1997)  and  Hansen  \&  Phinney  (1998)  also
presented cooling models for  low-mass He-core white dwarfs.  However,
because  these  studies  neglected  the evolutionary  history  of  the
progenitor stars, the role of hydrogen flashes and of residual nuclear
burning were not adequately addressed.  Hansen \& Phinney (1998) found
that thin envelopes  allow He-core white dwarfs to  cool rapidly while
for white dwarfs with  thick envelopes residual hydrogen burning slows
down  the  cooling.  The  importance  of  including  the evolution  of
progenitor  stars to properly  address the  role of  residual hydrogen
burning during the  final cooling branch was pointed  out by Driebe et
al.  (1998) and  Sarna et al.  (2000), who  reported the occurrence of
several  hydrogen  shell  flashes   during  the  cooling  phase.  Most
importantly, the  evolutionary calculations of  these authors resulted
in models with thick envelopes  in all the cases, including those that
experience thermonuclear  flashes and, hence, they  derived long white
dwarf  cooling  timescales  due  to residual  hydrogen  burning.   The
resulting  ages  are  thus  much  larger  than  those  resulting  from
considering  only the  decrease of  the thermal  content of  the ions.
More recently, Nelson  et al. (2004) also found  that hydrogen burning
strongly delays the cooling process, specially for very low-mass white
dwarfs.   They found flashes  to take  place for  masses in  the range
$0.21  \,  M_{\sun}\la M_{\rm  WD}  \la  0.28  \, M_{\sun}$,  in  good
agreement with the predictions of Driebe et al. (1998).

The studies  mentioned above show  that residual hydrogen  burning may
constitute the main  energy source for He-core white  dwarfs, even for
massive ones, a key aspect  in the interpretation of the observational
characteristics of these stars (Driebe at al. 1998). However, the role
of hydrogen  burning may be  strongly mitigated by  element diffusion.
Indeed, Iben \& Tutukov (1986) showed that chemical diffusion leads to
additional CNO flashes related  to diffusion of hydrogen inward, where
hotter  material  is  present.    These  authors  concluded  that  the
occurrence  of  similar  flashes  would  have the  effect  of  further
reducing the H-rich envelope, becoming eventually too small to sustain
any further  nuclear burning. The  importance of element  diffusion in
inducing  additional CNO flashes  was demonstrated  by Althaus  et al.
(2001a, 2001b), who found  that element diffusion strongly affects the
structure and cooling history  of He-core white dwarfs.  Indeed, these
authors found that diffusion-induced  CNO flashes yield small hydrogen
envelopes and, thus, at late  stages the evolution is fast. Althaus et
al. (2001a, 2001b)  found that diffusion is a  key physical ingredient
in explaining  the age dichotomy suggested by  observations of He-core
white dwarfs which are companions to millisecond pulsars (Bassa et al.
2003; Bassa  2006).  In addition,  Althaus et al.  (2001a)  found that
the mass range  for the occurrence of CNO  flashes depends strongly on
the inclusion of diffusion processes in evolutionary calculations.
 
The importance  of diffusion in He-core white  dwarfs with progenitors
having supersolar metallicities  has not been addressed so  far in the
literature. However, this is  an important issue that deserves further
study since it has relevant consequences for the age determinations of
stellar clusters.  The evolutionary  calculations to be presented here
together  with those  we  presented  in Serenelli  et  al. (2002)  for
He-core white dwarfs of low-metallicity progenitors constitute a solid
and consistent frame for the interpretation of observations of He-core
white dwarfs in stellar clusters having a wide range of metallicities.

\section{Computational details}

As previously stated, one of the aims of this work is to provide a set
of white dwarf cooling  tracks (including ages, colors and magnitudes)
appropriate for  studying low-mass He-core white  dwarfs in metal-rich
environments.   This grid  of models  covers a  wide range  of stellar
masses as detailed in Table 1.  Three supersolar metallicities for the
progenitor   star   have   been   considered:  $Z=$0.03,   0.04,   and
0.05. Specifically, we  have followed the evolution of  30 white dwarf
sequences from the  end of the mass-loss episode  during the pre-white
dwarf evolution down to very low surface luminosities.  The initial He
content at the main  sequence is given by $Y= 0.23 +  2.4 Z$, which is
consistent with  present determinations  of the chemical  evolution of
the Galaxy  (Flynn 2004; Casagrande  et al.  2007). Thus,  the initial
compositions of our sequences are, respectively, $(Y,Z)=(0.302,0.03)$,
$(Y,Z)=(0.326,0.04)$, and $(Y,Z)=(0.35,0.05)$.  To study the evolution
of He-core white  dwarfs with masses larger than  the helium core mass
at the helium flash, we  have also computed some white dwarf sequences
artificially suppressing  the occurrence of  the core helium  flash at
the tip of  the RGB.  Additional sequences without  diffusion and also
without nuclear burning  have also been computed to  assess the impact
of  nuclear  burning  and  diffusion  processes  on  the  evolutionary
properties of white dwarfs.  This allows to explore the role played by
diffusion in  the occurrence of  additional hydrogen shell  flashes in
He-core white dwarfs and,  more importantly, to investigate whether or
not the hydrogen envelope mass can be considerably reduced by enhanced
hydrogen burning during these flash episodes.  The evolutionary stages
prior  to  the formation  of  the white  dwarf  are  accounted for  by
computing the evolution of a $1  \, M_{\sun}$ model star for the three
mentioned metallicities from the main  sequence to the red giant stage
(see later in this section). In  what follows, we comment on the input
physics considered  in our calculations,  the initial models,  and our
atmospheric treatment.

\begin{table}
\begin{center}
\caption{Selected properties of our He-core white dwarf   sequences at
         the  point of maximum  effective temperature:  metallicity of
         progenitor star, stellar  mass, mass of hydrogen in the outer 
         layers, $M_{\rm H}$, and surface hydrogen abundance by mass.}
\begin{tabular}{@{}cccc}
\hline
\hline
$Z$ &$M/M_{\sun}$ & $M_{\rm H}/M_{\sun} [10^{-3}]$ & $X_{\rm H}$\\
\hline
0.03  &  0.220           &  2.07  &  0.490 \\
      &  0.250           &  1.51  &  0.531 \\ 
      &  0.303           &  0.98  &  0.658 \\
      &  0.358           &  0.56  &  0.658 \\
      &  0.400           &  0.40  &  0.658 \\
      &  0.452           &  0.27  &  0.658 \\
      &  0.513$^{\dag}$  &  0.20  &  0.658 \\
      &  0.522$^{\dag}$  &  0.17  &  0.658 \\
\hline
0.04  &  0.212           &  1.96  &  0.464 \\
      &  0.272           &  1.17  &  0.622 \\
      &  0.283           &  1.04  &  0.622 \\
      &  0.289           &  0.98  &  0.622 \\
      &  0.299           &  0.88  &  0.622 \\
      &  0.306           &  0.81  &  0.622 \\
      &  0.319           &  0.72  &  0.621 \\
      &  0.337           &  0.60  &  0.622 \\
      &  0.365           &  0.48  &  0.625 \\
      &  0.403           &  0.35  &  0.623 \\
      &  0.445           &  0.26  &  0.617 \\
\hline
0.05  &  0.207           &  2.26  &  0.430 \\
      &  0.273           &  1.06  &  0.588 \\
      &  0.288           &  0.91  &  0.588 \\
      &  0.295           &  0.84  &  0.588 \\
      &  0.301           &  0.78  &  0.588 \\
      &  0.311           &  0.72  &  0.588 \\ 
      &  0.323           &  0.63  &  0.588 \\
      &  0.348           &  0.51  &  0.588 \\
      &  0.368           &  0.43  &  0.585 \\
      &  0.401           &  0.34  &  0.588 \\
      &  0.445           &  0.25  &  0.588 \\
\hline
\hline
\end{tabular}
\end{center}
{\footnotesize $^{\dag}$These sequences  were obtained by artificially
               suppressing the core helium flash at the RGB.  They are
               discussed in Sect. 4.3.}
\end{table}

\subsection{Evolutionary code and input physics}

The  calculations presented  in  this  work have  been  done with  the
stellar evolutionary  code {\tt LPCODE}  --- see Althaus  et al.(2005)
and  references  therein.  This  code  has  been  previously used  for
studying  the  formation and  evolution  of  H-deficient white  dwarfs
(Althaus et al.   2005; Miller Bertolami \& Althaus  2006) and extreme
horizontal branch stars (Miller Bertolami  et al.  2008).  The code is
based  on  a  detailed  description  of the  main  physical  processes
involved in the formation of white dwarfs through late thermal pulses,
particularly non-instantaneous  mixing --- see Althaus  et al.  (2005)
for details.   The standard mixing length theory  for convection (with
the free parameter $\alpha=1.6$) has been adopted.  With regard to the
microphysics  considered  in  the  present calculations,  we  employed
radiative opacities  from OPAL for arbitrary metallicity  in the range
from  0  to  0.1  (Iglesias  \&  Rogers  1996),  supplemented  at  low
temperatures   with  the  Alexander   \&  Ferguson   (1996)  molecular
opacities.  Neutrino emission rates for pair, photo and bremsstrahlung
processes were taken  from Itoh et al.  (1996).   For plasma processes
we  included   the  treatment  presented  in  Haft   et  al.   (1994).
Conductive opacities are from Cassisi et al.  (2007), which covers the
whole  regime where electron  conduction is  relevant.  For  the white
dwarf regime we employed an updated version of the Magni \& Mazzitelli
(1979) equation of  state.  The nuclear network takes  into account 16
elements and 34 nuclear reactions  for pp chains, CNO bi-cycle, helium
burning  and   carbon  ignition,  and  are  detailed   in  Althaus  et
al. (2005).

All  our  sequences  have  been  computed in  a  self-consistent  way,
including the evolution of  the chemical abundance distribution caused
by  element diffusion  processes and  burning during  the  whole white
dwarf  stage.   It  is  worth  mentioning that  the  opacity  is  also
calculated  self-consistently using the  predictions of  diffusion for
the heavy element  composition.  In particular, for the  case in which
diffusion is considered, metallicity  is taken as twice the abundances
of  CNO elements.   We mention  that time-dependent  diffusion  due to
gravitational settling, and chemical  and thermal diffusion of nuclear
species has been fully taken into account following the multicomponent
gas  treatment   of  Burgers   (1969).   The  following   species  are
considered:  $^{1}$H,  $^{3}$He,   $^{4}$He,  $^{12}$C,  $^{14}$N  and
$^{16}$O.  The  resulting flow equations  for these species  have been
integrated by using a semi-implict, finite-difference scheme.  Details
are given  in Althaus  et al.  (2001a,  2005).  Abundance  changes are
computed according to element  diffusion and then to nuclear reactions
and convective mixing. This detailed treatment of abundance changes by
different processes  during the white  dwarf regime constitutes  a key
aspect in the evaluation of the importance of residual nuclear burning
for the cooling of low-mass white dwarfs.

\subsection{Initial models}

The existence  of a population of  single He-core white  dwarfs is not
yet discarded  (nor confirmed).  Recent theoretical  evidence based on
energetic  considerations  suggests that  these  white  dwarfs may  be
descendants of single  supersolar metallicity progenitors with initial
mass $M  \leq 1 \,  M_{\sun}$ (Meng et  al.  2008).  Hence,  to derive
starting configurations for our more massive He-core cooling sequences
consistent with  the evolutionary history  of the progenitor  star, we
have  simply  removed  mass  from  a  $1 \,  M_{\sun}$  model  at  the
appropriate stages of  its evolution (Iben \& Tutukov  1986; Driebe et
al.   1998).  Specifically, we  forced the  progenitor star  to depart
from the red giant phase just before the occurrence of the helium core
flash.  This is done for the three metallicities considered here.  The
resulting  final stellar  masses for  each metallicity  are  listed in
Table 1  together with the total  amount of hydrogen  contained in the
envelope,  $M_{\rm  H}$, and  the  surface  abundance  of hydrogen  at
maximum $T_{\rm eff}$  at the beginning of the  cooling branch.  It is
important to  mention at  this point that  as far as  the evolutionary
properties  of our  He-core  white dwarfs  are  concerned, a  detailed
knowledge of the progenitor evolution, in particular the initial mass,
is not a relevant issue (Driebe et al. 1998).

A look at Table  1 shows that the larger the mass  of the remnant, the
smaller the  residual mass of the  hydrogen in the outer  layers, in a
way similar  to what is  found for post-AGB  remnants.  As it  will be
shown later,  this hydrogen envelope will be  responsible for residual
hydrogen  burning, thus  markedly lengthening  the cooling  times.  In
this  sense, the  values  of $M_{\rm  H}$  listed in  Table  1 can  be
considered  as an upper  limit to  the expected  evolutionary hydrogen
envelope  with which  a  given  He-core white  dwarf  enters into  its
cooling track.  We  also note from Table 1 that  the values of $M_{\rm
H}$ depend  on the metallicity of the  progenitor star.  Specifically,
$M_{\rm  H}$   decreases  as  the   progenitor  metallicity  increases
(Castellani et al.  1994).  Because of this, the values of $M_{\rm H}$
derived here are  somewhat smaller than those quoted  by Driebe et al.
(1998) in their  table 2 for solar  metallicity progenitors.  Finally,
note that for stellar masses  below $M \approx 0.25 \, M_{\sun}$, mass
loss  during the  RGB evolution  has uncovered  layers  where hydrogen
burning took  place in prior  stages, giving rise to  pre-white dwarfs
with He-enriched outer layers.

We have  applied the same procedure to  obtain starting configurations
in the case of He-core  white dwarfs with smaller masses.  These white
dwarfs  are probably  the  result  of mass  transfer  in close  binary
systems  (Sarna et  al. 2000),  since exceedingly  high ages  would be
needed  to produce  very  low-mass white  dwarfs  through single  star
evolution.  However, we have  not modeled the binary evolution leading
to  the formation  of these  stars, as  it was  done in  Panei  et al.
(2007).   It  is worth  mentioning  in  this  context that  after  the
termination of the mass-loss episodes, the subsequent evolution of the
model does not  depend on the details of how most  of the envelope was
lost.   In particular,  the mechanical  and thermal  structure  of the
models  are  consistent  with  the  predictions  of  binary  evolution
calculations (Driebe et al. 1998).

Finally,  for  He-core  white  dwarfs  with  masses  larger  than  the
canonical  value  of  the  core  mass  at the  helium  flash  we  have
artificially suppressed the occurrence of the core helium flash at the
tip of  the RGB.  In  this way, we  have derived He-core  sequences of
stellar masses  of $0.513$  and $0.522 \,  M_{\sun}$.  Whether  or not
this is a realistic procedure  deserves further study, which is beyond
the scope  of this  paper. {\bf We  mention that because  of numerical
difficulties,  white  dwarf models  with  higher  He  cores have  been
computed only for the case of $Z=0.03$}.

\subsection{Model atmosphere}

For  all  the  evolutionary  sequences  in which  diffusion  has  been
considered, we compute accurate colors  and magnitudes on the basis of
improved LTE model  atmospheres. The numerical code used  is a new and
updated version of  that described in Rohrmann et  al.  (2002). Models
are    computed   assuming   hydrostatic    and   radiative-convective
equilibrium. Convective transport present in the cooler atmospheres is
treated   within   the   usual   mixing-length   approximation.    The
microphysics  included in  the model  atmospheres  comprises non-ideal
effects in the gas equation of state and chemical equilibrium based on
the occupation  probability formalism as described in  Rohrmann et al.
(2002). The  code includes H,  H$_2$, H$^+$, H$^-$,  H$_2^+$, H$_3^+$,
He, He$^-$, He$^+$, He$^{2+}$, He$_2^+$, HeH$^+$, and e$^-$. The level
occupation  probabilities are  self-consistently  incorporated in  the
calculation  of the  line and  continuum  opacities. Collision-induced
absorptions due to H$_2$-H$_2$, H$_2$-He and H-He pairs are also taken
into account (Rohrmann et al. 2002).

For the purpose of the  present work, the model atmospheres explicitly
include   Ly$\alpha$   quasi-molecular   opacity  according   to   the
approximation  used  by Kowalski  \&  Saumon (2006).   Quasi-molecular
absorption   results   from  perturbations   of   hydrogen  atoms   by
interactions  with  other particles,  mainly  H  and  H$_2$. Here,  we
consider   extreme   pressure-broadening   of  the   line   transition
H($n=1$)$\rightarrow$H($n=2$) due to  H-H and H-H$_2$ collisions, with
the  red wing  extending  far  into the  optical  region.  A  detailed
description of  these collisional line-broadening  evaluations will be
presented  in a  forthcoming paper  (Rohrmann et  al.  2009).   On the
basis of the approximations outlined  in Kowalski \& Saumon (2006), we
evaluate  the red  wing  absorption within  the quasi-static  approach
using  theoretical molecular  potentials to  describe  the interaction
between  the  radiator  and  the  perturber.   We  also  consider  the
variation   of  the   electric-dipole  transition   moment   with  the
interparticle   distance.    The   H$_3$   energy-potential   surfaces
contributing  to collisions  H-H$_2$  were taken  from the  analytical
representations of  Varandas et al. (1987), and  the dipole transition
moments were calculated from  Petsalakis et al. (1988).  Broadening of
Ly$\alpha$  line by  H-H collisions  plays  a minor  role compared  to
H-H$_2$ encounters. The  potential interactions for H-H configurations
were  taken  from  Kolos  \&  Wolniewicz  (1965)  and  the  transition
probability was assumed constant in this case.  The main effect of the
Ly$\alpha$  quasi-molecular opacity  is a  reduction of  the predicted
flux at  wavelength smaller than  $5000$ \AA\ for white  dwarfs cooler
than $T_{\rm eff} \approx 6000$~K.

\begin{figure}
\centering
\includegraphics[clip,width=220pt ]{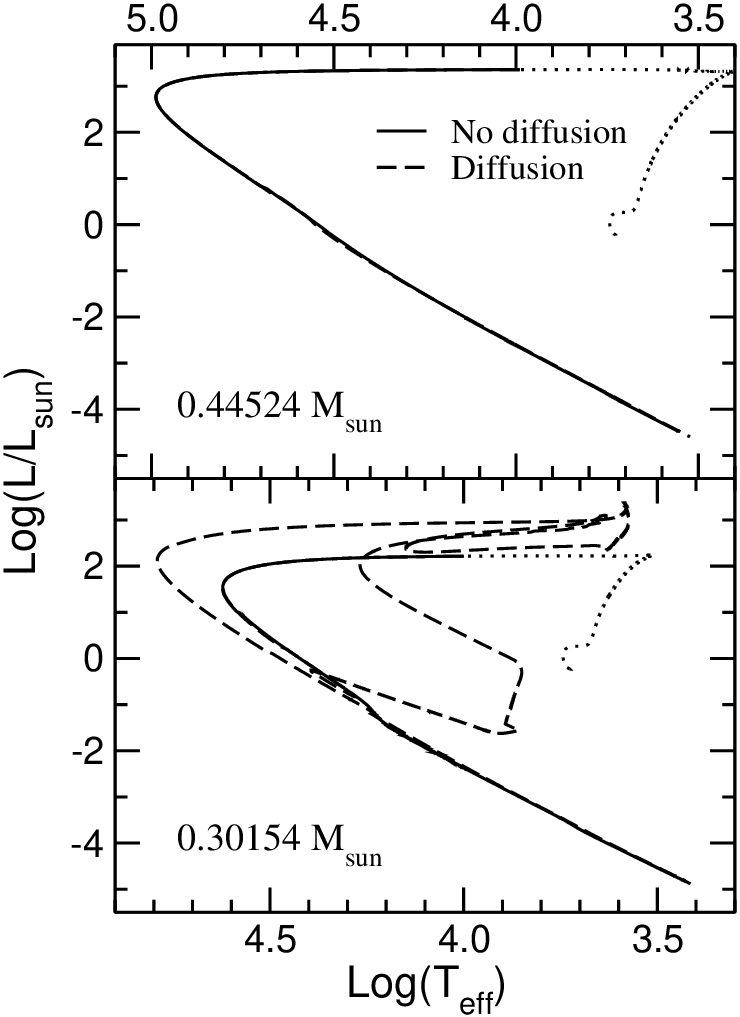}  
\caption {Hertzsprung-Russell diagram for selected He-core white dwarf
          evolutionary sequences  obtained from a  $Z=0.05$ progenitor
          star.   Dashed (solid)  lines  correspond to  the case  when
          element diffusion  is (not)  taken into account.  The dotted
          lines show  the evolution  of the $1\,  M_{\sun}$ progenitor
          star.}
\label{hr}
\end{figure}

\begin{figure}
\centering
\includegraphics[clip,width=220pt]{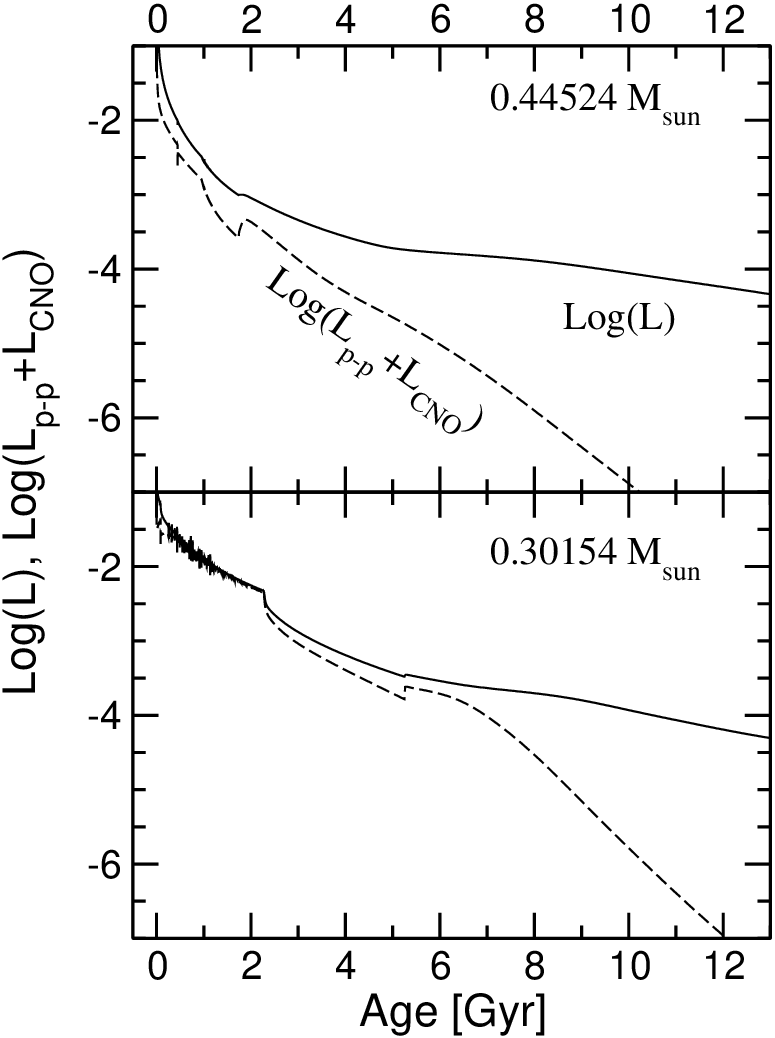}
\caption{Surface luminosity  (solid lines) and  nuclear luminosity due
         to hydrogen burning  (dashed lines) as a function  of age for
         two  selected  He-core  cooling  sequences.  The  panels  are
         labelled with the  corresponding stellar mass.  The sequences
         correspond  to  the  case   in  which  we  consider  $Z=0.05$
         progenitors.  Diffusion processes are not considered.}
\label{luminodif}
\end{figure}

\section{Evolutionary results}

In  this  section  we  describe  the evolutionary  properties  of  our
sequences,  placing strong emphasis  on the  role of  residual nuclear
burning during the  white dwarf stage. We explore  the implications of
element diffusion  and we investigate  the evolution of  He-core white
dwarfs  with stellar  masses larger  than the  canonical value  of the
helium core mass at the helium flash.

\begin{figure*}
\centering
\includegraphics[clip,width=350pt]{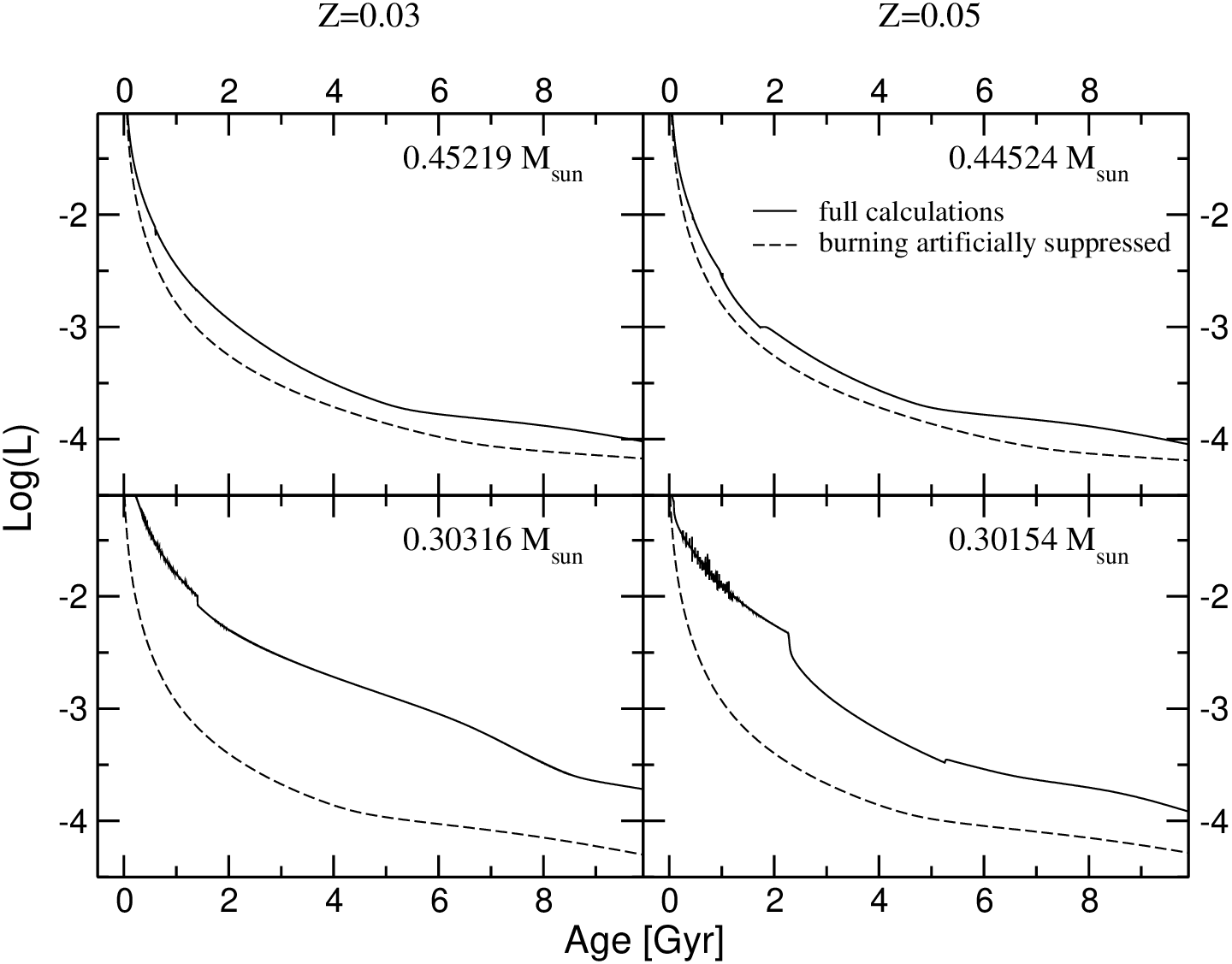}
\caption{Surface luminosity as a  function of age for selected He-core
         sequences.   Each panel  is labelled  with  the corresponding
         stellar mass. Panels are also labelled with the corresponding
         metallicity of  the progenitor  stars.  The solid  lines show
         the  results of the  self-consistent calculations,  while the
         dashed lines  display the evolution when  hydrogen burning is
         artificially   suppressed.     Element   diffusion   is   not
         considered.}
\label{luminuc_z03-05}
\end{figure*}

\subsection{Evolutionary sequences without element diffusion}

We begin  by examining the results corresponding  to the approximation
usually adopted in  most studies of these stars  in which diffusion is
not  considered.   As  discussed  previously, despite  the  fact  that
chemical  diffusion  operating at  the  base  of  the H-rich  envelope
appears to be  an attractive hypothesis to match  the pulsar spin-down
ages  and  the low-mass  white  dwarf  cooling  ages (Althaus  et  al.
2001a), its  validity is still  not well established,  since diffusion
could be  mitigated and even  suppressed by other  physical processes,
like  strong  magnetic  fields.    Exploring  this  possibility  is  a
difficult task that would carry us too far afield. However, we believe
to  be  worthwhile to  explore  the  extreme  situation in  which  the
evolution  of  He-core  white  dwarfs  is not  affected  by  diffusion
processes. In  fact, the evolutionary  sequences computed disregarding
diffusion allow  to establish an upper  limit to the  role of residual
hydrogen burning.

We  concentrate on  the  sequences with  $M=0.44524$  and $0.30154  \,
M_{\sun}$ with  progenitors of metallicity $Z=0.05$. We  will refer to
these  sequences  as  the  standard  sequences.   Specifically,  these
sequences have been  calculated neglecting both gravitational settling
and chemical diffusion and assuming  a metallicity of $Z=0$ during the
white dwarf regime.  The  Hertzsprung-Russell diagram for our selected
standard  sequences is  displayed  in Fig.  \ref{hr}.  The absence  of
hydrogen  shell  flashes in  the  selected  standard  sequences is  an
expected  feature and  it is  in  agreement with  previous studies  of
low-mass white  dwarfs.  We find  that hydrogen shell  burning becomes
unstable only for stellar masses below $\approx 0.26 \, M_{\sun}$.

We  find that  residual hydrogen  burning via  the pp  chain reactions
considerably  slows down  the cooling  rate of  evolved  He-core white
dwarfs resulting from metal-rich progenitors, in a similar way to what
it has been  reported for low-mass He-core white  dwarfs stemming from
progenitors with smaller metal content (Castellani et al. 1994; Driebe
et al. 1998;  Sarna et al. 2000).  The  importance of residual nuclear
burning as a  source of energy is made  clear in Fig. \ref{luminodif},
which shows the temporal  evolution of the hydrogen burning luminosity
and the surface  luminosity for the 0.44524 and  $0.30154 \, M_{\sun}$
sequences.  We  have chosen to  show the results corresponding  to the
case in which  a $Z=0.05$ progenitor was adopted.   Only the evolution
on  the final  cooling phase  is depicted  in the  figure.   Note that
residual hydrogen burning --- for  the stages shown in this figure the
pp  chain reactions are  dominant ---  appreciably contributes  to the
surface  luminosity  of the  star  even  at  very advanced  stages  of
evolution. This is  so despite the fact that the  mass of the hydrogen
retained in the  surface layers at the beginning  of the cooling phase
is markedly  smaller in metal-rich precursor  stars.  The contribution
of residual nuclear  burning to the energy budget  of the star becomes
more and more  dominat as the stellar mass is  decreased. We find that
nuclear burning  is also dominant  at advanced evolutionary  stages of
He-core sequences that experience hydrogen shell flash episodes, i.e.,
for stellar masses below $\approx  0.26 \, M_{\sun}$. Finally, we find
that  the contribution of  residual nuclear  burning increases  as the
metallicity of the progenitor star decreases (see below).

\begin{figure}
\centering
\includegraphics[clip,width=220pt]{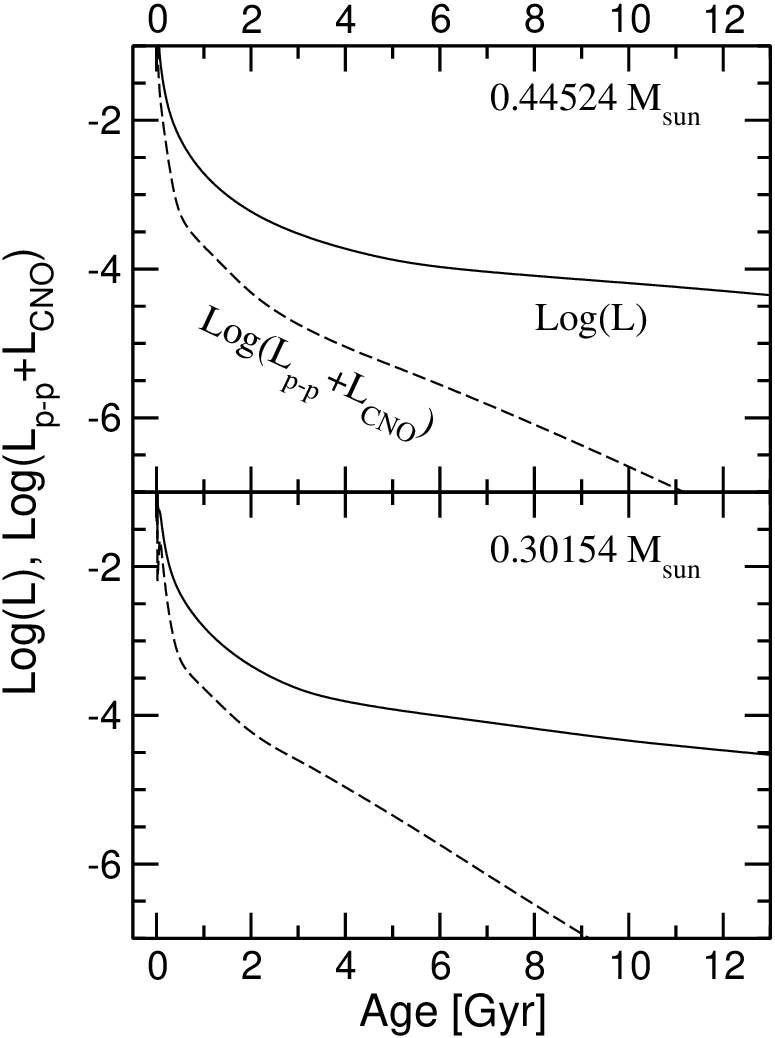}
\caption{Same  as  Fig.  \ref{luminodif}   but  for  the case in which 
         diffusion processes are  fully taken into account.}
\label{lumi-dif-z05}
\end{figure}

\begin{figure*}
\centering
\includegraphics[clip,width=350pt]{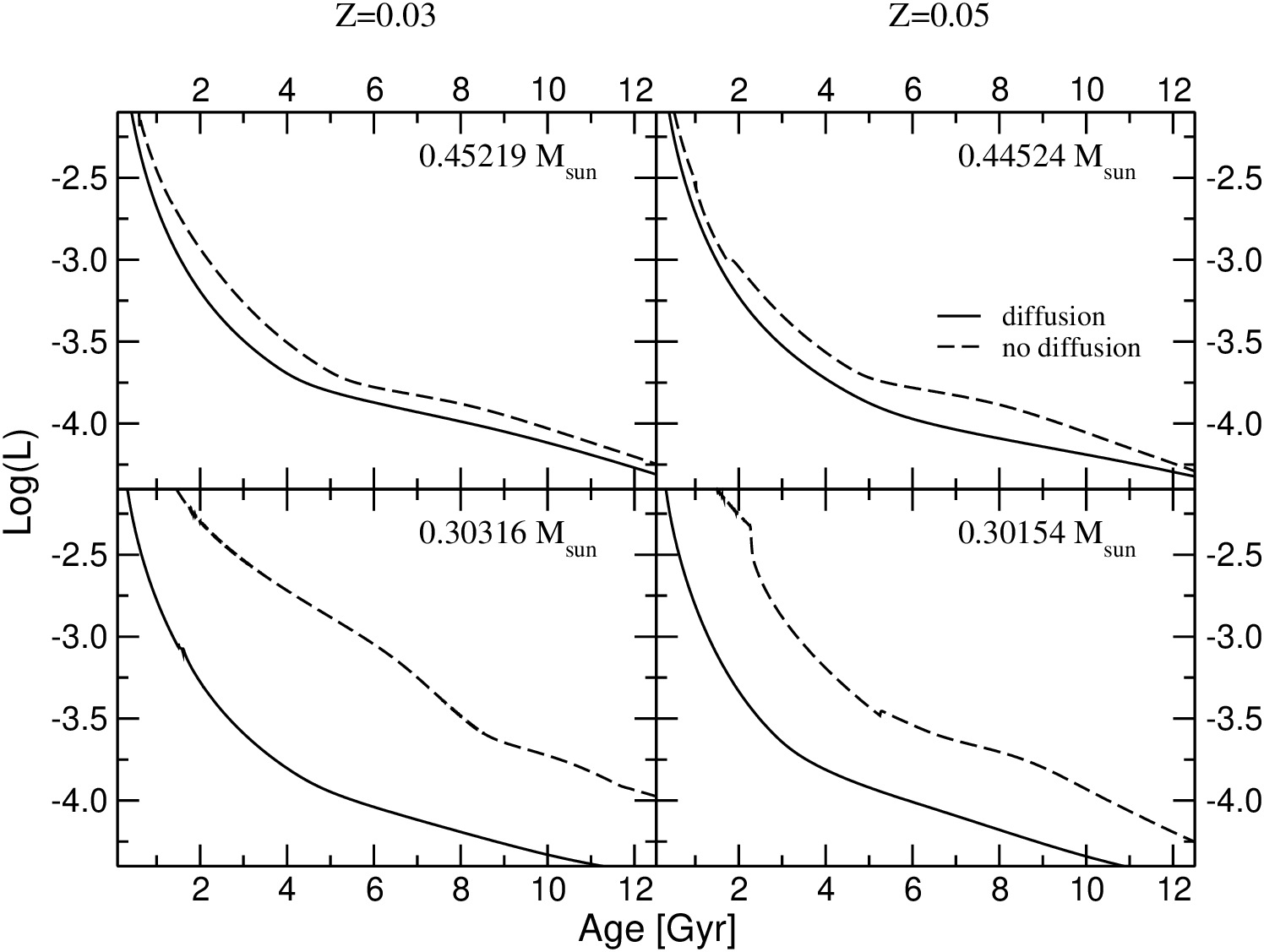}
\caption{Surface luminosity as a function of age for selected  He-core
         sequences as labelled by  their stellar mass for $Z=0.05$ and
         $Z=0.03$     progenitors,    right     and     left    panels
         respectively.   Solid  (dashed)   lines  correspond   to  the
         situation when element diffusion is (not) considered.}
\label{comparo}
\end{figure*}

The impact  of residual  nuclear burning on  the evolution  of He-core
white dwarfs  becomes clear by  inspecting Fig.  \ref{luminuc_z03-05},
which displays  the temporal evolution  of the surface  luminosity for
selected   He-core   sequences   resulting   from   progenitors   with
metallicities  $Z=0.05$ and  $Z=0.03$.   In this  figure, solid  lines
correspond to the results of the full evolutionary calculations, while
the  dashed lines illustrate  the situation  when hydrogen  burning is
artificially   suppressed.   Note   that  residual   hydrogen  burning
considerably  influences  the age  of  the  white  dwarf. Indeed,  the
evolution is delayed to very  long ages by the active hydrogen burning
zone.  Clearly,  when  nuclear  burning  is  neglected  a  substantial
underestimate of the cooling age  of the white dwarf is obtained.  The
magnitude of  the delays introduced  by nuclear burning is  larger for
decresing stellar masses and  metallicities.  In most cases, evolution
is essentially dictated  by nuclear burning, giving rise  to very long
cooling ages.   For the $0.44524 \, M_{\sun}$  sequence with $Z=0.05$,
the  magnitude  of   the  delays  amounts  to  0.4   and  2.7  Gyr  at
$\log(L/L_{\sun})=-3$ and  $-4$, respectively. These  delays amount to
2.2 and  5.3 Gyr for  the $0.30154 \,  M_{\sun}$ sequence at  the same
luminosities. The  effect of the  metallicity is also  quite apparent,
increasing the cooling ages. For instance, for $Z=0.03$, the delays in
the  cooling times  amount to  0.8  (4.6) and  3.5 (7.3)  Gyr for  the
0.45219   (0.30316)   $M_{\sun}$  cooling   sequences   at  the   same
luminosities.  In summary,  it is clear that the  evolution of He-core
white  dwarfs   resulting  from  progenitors   stars  with  supersolar
metallicity is considerably affected  by nuclear burning. This is also
true  in the  case of  massive He-core  white dwarfs.   In particular,
residual pp hydrogen burning strongly slows the evolutionary rate down
to very low luminosities, a fact that must be considered when attempts
are made to date metal-rich stellar populations using these stars.

\subsection{The effect of element diffusion}

We find that diffusion processes acting during the white dwarf cooling
track  strongly modify  the shape  of the  chemical  profiles, causing
hydrogen to float  to the surface, and helium  and heavier elements to
sink  down, as  expected. As  a result,  the pre-white  dwarf  with an
initially H-  and He-rich  envelope turns into  an object with  a pure
hydrogen envelope, that is, a white  dwarf of the spectral type DA. We
also  find  that  the  tail  of the  innermost  hydrogen  distribution
chemically diffuses  inwards into hotter layers  as evolution proceeds
on  the cooling track.   At high  effective temperatures,  this effect
favors   the   occurrence  of   CNO   thermonuclear  flashes.    These
diffusion-induced hydrogen  flashes are responsible for  the fact that
the hydrogen envelope with which the white dwarf enters into its final
cooling track becomes markedly thinner than in the case when diffusion
is  neglected.  These  results  are  very similar  to  those found  by
Althaus et  al (2001a, 2001b) and  Panei et al. (2007)  about the role
that  element  diffusion plays  in  inducing additional  thermonuclear
flashes,  and being  responsible  for the  occurrence  of thin  H-rich
envelopes  and a fast  evolution at  late stages  of the  evolution of
He-core white dwarfs with  solar metallicity progenitors.  The results
presented  here show  that the  same occurs  for He-core  white dwarfs
resulting from  metal-rich progenitors.  It is also  worth noting that
the occurrence of diffusion  modifies the threshold for the occurrence
of hydrogen shell  flashes.  In the case of  He-core white dwarfs with
$Z=0.05$ progenitors,  we find that  this threshold is about  $0.37 \,
M_{\sun}$, while  for solar metallicity progenitors  all the sequences
exhibit flashes.   The He-core  cooling sequences with  stellar masses
smaller  than  this  do  experience thermonuclear  flashes  (see  Fig.
\ref{hr} for the case of the $0.30154 \, M_{\sun}$ sequence).  Because
of the larger abundance of hydrogen carried by chemical diffusion into
the burning region, this threshold value becomes larger for decreasing
metallicities of the parent star.

\begin{figure}
\centering
\includegraphics[clip,width=220pt]{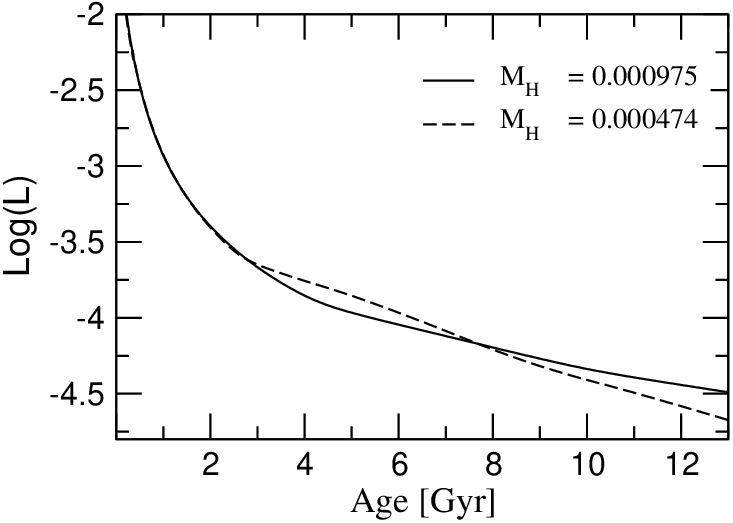}
\caption{Surface luminosity as a function of age for our He-core white
         dwarf  sequence  of  $0.30316  \, M_{\sun}$  derived  from  a
         $Z=0.03$  progenitor  and  with different  hydrogen  envelope
         masses. In both cases,  element diffusion has been considered
         and   residual   nuclear   burning  has   been   artificially
         suppressed.  The  mass of the hydrogen envelopes  is given in
         solar masses.}
\label{opacidad}
\end{figure}

As  mentioned, as  a  result of  the  occurrence of  diffusion-induced
flashes,  no residual  nuclear burning  is expected  during  the final
cooling branch.   This is made clear in  Fig. \ref{lumi-dif-z05} which
displays the  temporal evolution  of the hydrogen-burning  and surface
luminosities  during the  final cooling  branch for  the  same stellar
masses  shown in  Fig. \ref{luminodif}.   Note that  the  inclusion of
diffusion   predicts  residual   hydrogen  burning   to  be   a  minor
contribution to  the star luminosity. In  the case of  the $0.44542 \,
M_{\sun}$  sequence, which does  not suffer  from hydrogen  flashes, a
large fraction of the hydrogen  content is burnt during the hot stages
of the  evolution.  Thus, also in  the case of sequences  which do not
suffer from hydrogen flashes, the  white dwarf is ultimately left with
a thin hydrogen-rich envelope, when diffusion is considered.

\begin{figure}
\centering
\includegraphics[clip,width=220pt]{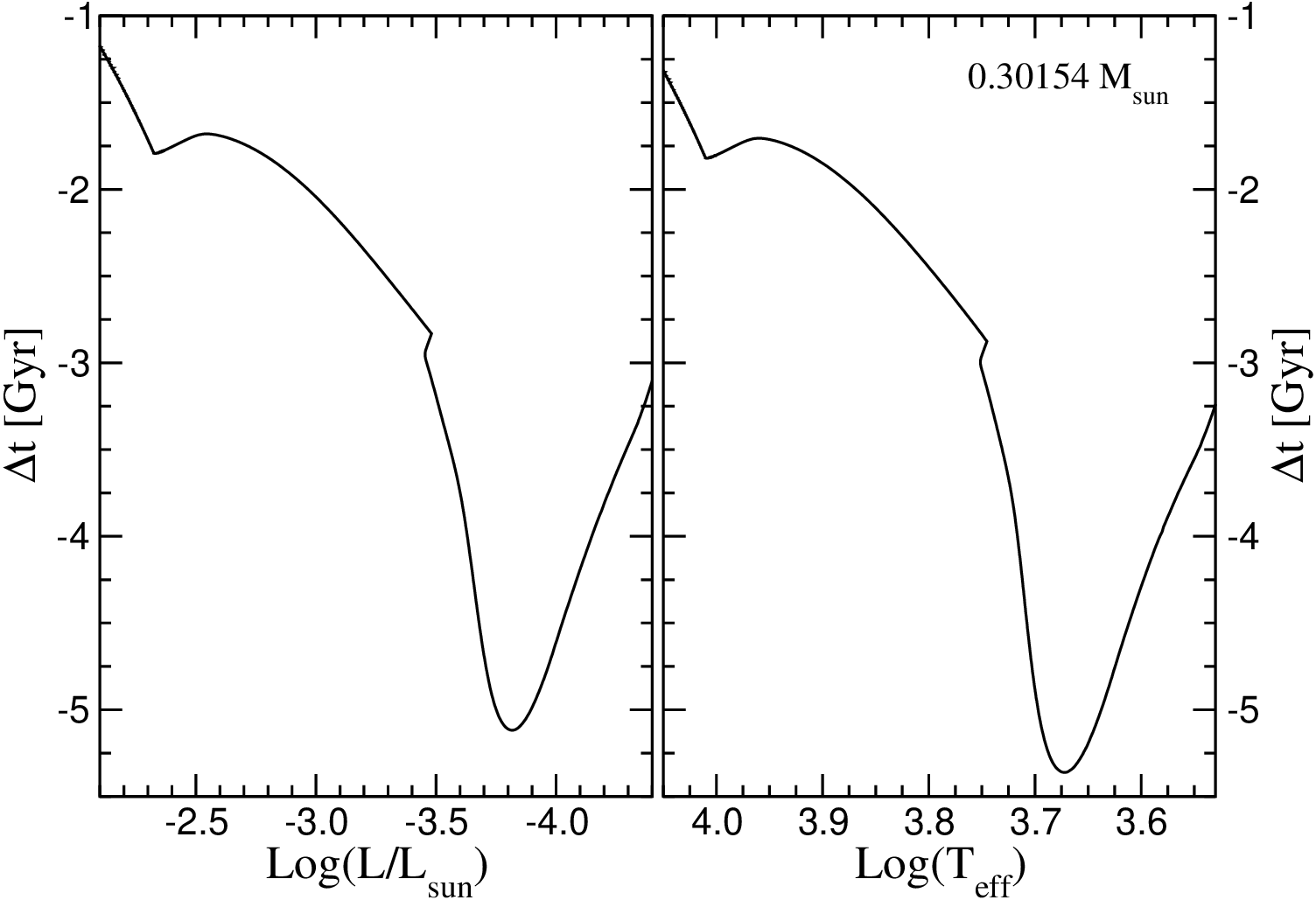}
\caption{Differences  in cooling  times  (in Gyr)  resulting from  the
         inclusion  of diffusion  corresponding to  the  sequence with
         $0.30154 \, M_{\sun}$ derived from a $Z=0.05$ progenitor.}
\label{delay}
\end{figure}

We expect  the evolutionary timescales  of He-core white dwarfs  to be
considerably affected  by the inclusion  of diffusion.  This  is borne
out in Fig. \ref{comparo}, which clearly shows the impact of diffusion
on  the temporal  evolution  of  the surface  luminosity  for the  two
stellar  masses we  analyzed previously.   Note that  the evolutionary
ages  are substantially  smaller when  diffusion is  considered.  This
reduction is  more noticeable for smaller metallicities  of the parent
star  and  stellar  masses.  {\bf  The differences  in  cooling  times
resulting   from  the   inclusion  of   diffusion  are   displayed  in
Fig. \ref{delay} for the $0.30154 \, M_{\sun}$ sequence. Note that the
delay in  cooling times  can reach  up to 5  Gyr (because  of residual
nuclear  burning) if diffusion  is not  considered}. We  conclude that
diffusion prevents hydrogen burning from being a main source of energy
for  most of  the evolution  of He-core  white dwarfs  with metal-rich
progenitors.  As a  result, white dwarfs obtain energy  from its relic
thermal  content, and  the cooling  ages become  notably  smaller when
compared  with  those obtained  in  the  case  in which  diffusion  is
neglected.   Indeed, as we  showed previously,  when diffusion  is not
considered, residual  hydrogen burning after  thermonuclear flashes is
the dominant energy source even at advanced stages of evolution.

\begin{figure}
\centering
\includegraphics[clip,width=220pt]{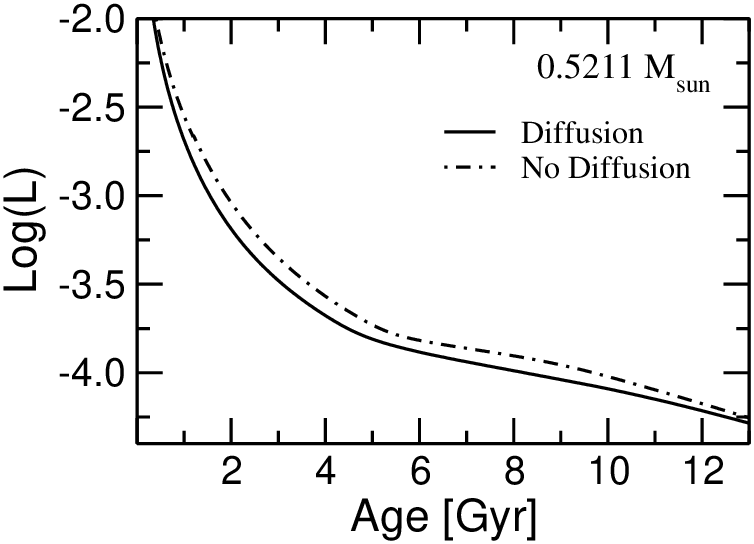}
\caption{Surface  luminosity  as  a  function  of  age for our massive 
         He-core white dwarf sequence  of $0.5211 \, M_{\sun}$ derived
         from a $Z=0.03$ progenitors with artificially suppressed core
         helium  flash. The  solid  (dashed) line  corresponds to  the
         situation in which element diffusion is (not) considered.}
\label{fig_masiva}
\end{figure}

{\bf Finally,  it should be  noted that there  is a dependence  of the
cooling  times on  the  thickness of  the  hydrogen-rich envelope.  To
assess this effect, we have  computed the evolution of the $0.30316 \,
M_{\sun}$  sequence for  two different  situations. In  particular, in
Fig. \ref{opacidad} we show as a solid line the sequence corresponding
to  the case of  a thick  hydrogen envelope.   This sequence  has been
derived by artificially suppressing the nuclear burning, thus avoiding
the occurrence of the flash and the formation of a thin envelope.  The
sequence depicted  with dashed line illustrates the  situation for the
case  of a  thin  envelope that  results  from the  occurrence of  the
hydrogen  flash. Both sequences,  though not  entirely self-consistent
with the assumptions of the modeling considered in this work, allow us
to  isolate  the opacity  effects  from  the  nuclear energy  effects.
Although the cooling times do  show a non-negligible dependence on the
mass of the  residual hydrogen envelope, we emphasize  that the effect
of  the  opacity  resulting  from  a different  layering  (Tassoul  et
al.  1990) is  much weaker  than the  effects resulting  from residual
hydrogen burning.}

\subsection{Massive helium-core white dwarfs}

The possibility that some He-core  white dwarfs in NGC 6791 could have
masses larger than the canonical value  of the core mass at the helium
flash ($\approx 0.45 \,  M_{\sun}$ for metal-rich, low-mass RGB stars)
has  been recently raised  by Bedin  et al.   (2008b) to  explain some
observed features of the RGB and  of the cooling sequence of NGC 6791.
We  have  extended  the  scope   of  the  present  work  by  computing
evolutionary  sequences for  He-core white  dwarfs with  masses larger
than $0.45 \,  M_{\sun}$.  To this end, as explained  in Sect. 3.2, we
have artificially  suppressed the occurrence of the  core helium flash
at the tip  of the RGB.  In this way, we  derived He-core sequences of
stellar masses of $0.513$ and $0.522 \, M_{\sun}$.

Evolutionary results for these  massive white dwarfs are summarized in
Fig.   \ref{fig_masiva}, which  displays the  surface luminosity  as a
function of age  for a He-core sequence of  $0.5211\, M_{\sun}$ with a
$Z=0.03$ progenitor.  The solid line shows the case in which diffusion
is considered,  while the  dashed line illustrates  the case  in which
diffusion is  neglected. Following the trend  we find for  the case of
the  more massive  He-core white  dwarfs discussed  previously,  it is
clear that  the age  of massive He-core  white dwarfs is  not markedly
influenced  by the  occurrence  of diffusion  processes.   This is  an
expected  result because  for such  massive He-core  white  dwarfs the
H-rich envelope  left after RGB evolution is  substantially smaller in
this case,  see Table  1, with the  consequence that  residual nuclear
burning is not  a main source of energy even in  the case that element
diffusion  is  neglected.  This  behavior  is  in  contrast  with  the
situation we  find in  less massive He-core  white dwarfs,  for which,
when  diffusion is  neglected,  nuclear burning  constitutes the  main
energy  source even at  advanced stages,  delaying their  evolution to
very long ages.

\section{Colors and magnitudes}

\begin{figure}
\centering
\includegraphics[clip,width=220pt]{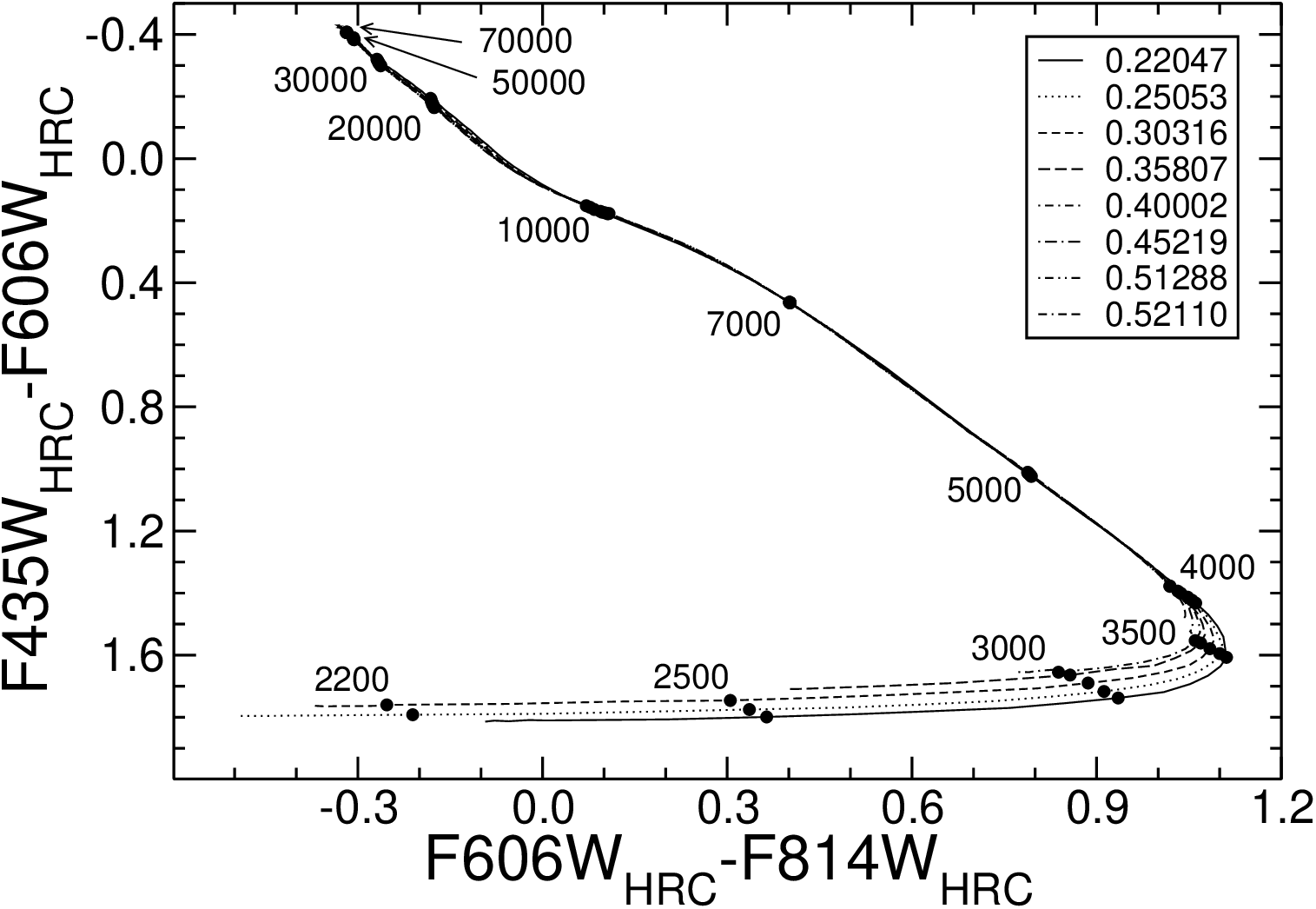}
\caption{F435W--F606W,   F606W--F814W  color-color  diagram   for  our
         He-core  white  dwarf  sequences with  $Z=0.03$  progenitors.
         Pure hydrogen composition  is assumed. Selected $T_{\rm eff}$
         values are labeled along the curves.}
\label{cc}
\end{figure}

\begin{figure}
\centering
\includegraphics[clip,width=220pt]{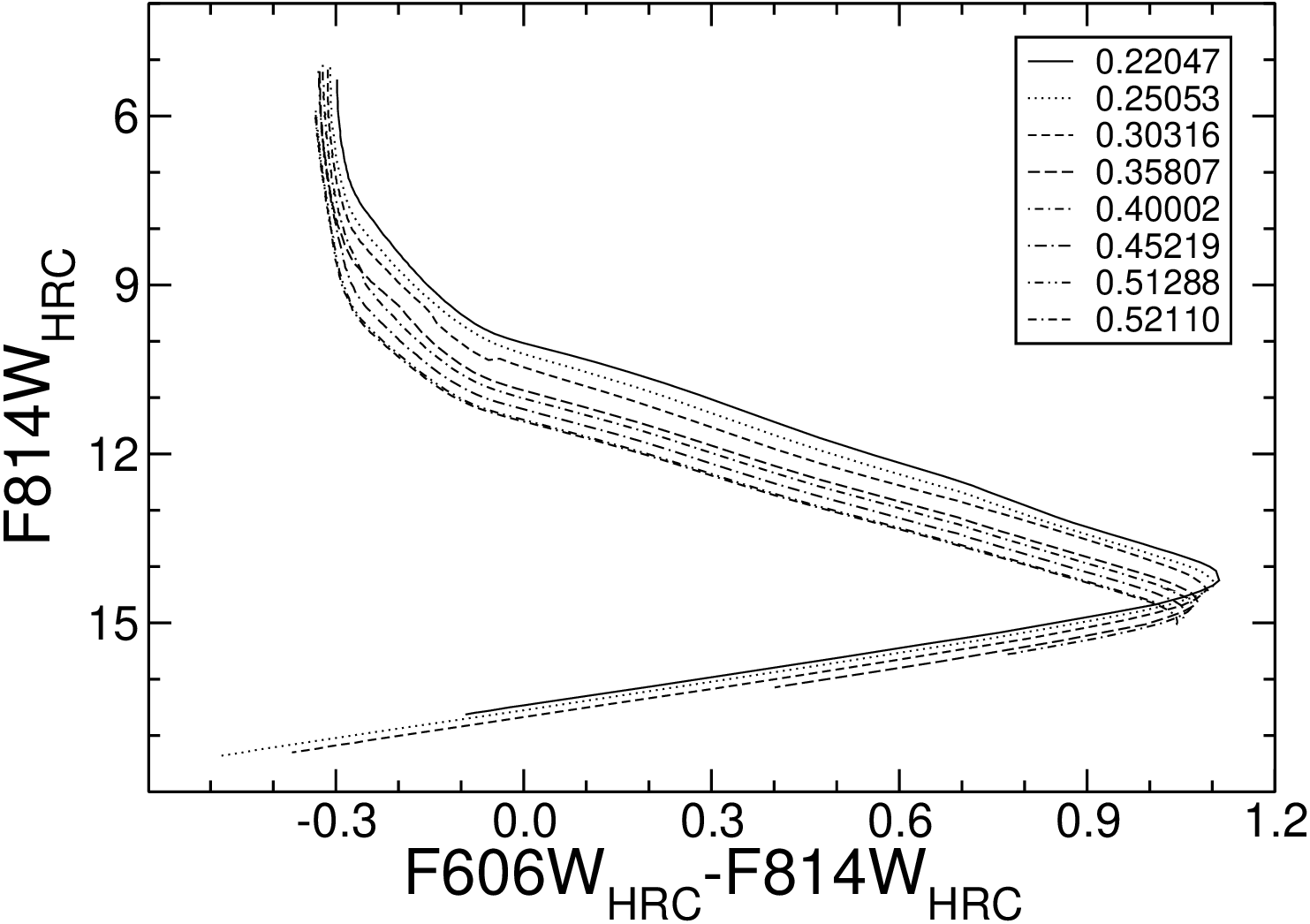}
\caption{F814W, F606W--F814W  color-magnitude diagram for  our He-core
         white  dwarf  sequences   with  $Z=0.03$  progenitors.   Pure
         hydrogen composition is assumed.}
\label{cmd}
\end{figure}

For all our  cooling sequences in which diffusion  has been considered
we  have computed  accurate colors  and  magnitudes based  on new  and
improved   non-gray  model   atmospheres   which  explicitly   include
Ly$\alpha$ quasi-molecular opacity according to the approximation used
by Kowalski \&  Saumon (2006).  The calculations have  been made for a
pure  hydrogen  composition and  for  the  HST  ACS filters  (Vega-mag
system) and $UBVRI$ photometry.   As an example of these calculations,
we  show in  Figs.  \ref{cc}  and \ref{cmd}  selected  color-color and
color-magnitude   diagrams  for  all   our  sequences   with  $Z=0.03$
progenitors.   In both diagrams,  all our  sequences exhibit  the well
known turn-off in  their colors at very low  effective temperature and
become   blue  with   further  evolution.    This  is   a   result  of
collision-induced absorption  from molecular hydrogen,  a process that
reduces the  infrared flux  and forces radiation  to emerge  at larger
frequencies.   Finally, we  mention that  the inclusion  of Ly$\alpha$
quasi-molecular opacity makes the  sequences redder, thus delaying the
turn  to the  blue (particularly  in the  $UBVRI$  photometric system,
which  we do  not show  here  for the  sake of  conciseness). This  is
because  the collisional  effect on  Ly$\alpha$ reduces  the predicted
flux at wavelengths smaller than $5000$ \AA\ for cool white dwarfs.

\section{Discussion and conclusions}

In this  paper we have studied  the evolution of  He-core white dwarfs
resulting  from  metal-rich  progenitors.   The  absence  of  detailed
calculations  for such  white  dwarfs,  as well  as  the discovery  of
He-core white dwarfs in the  metal-rich cluster NGC 6791 have prompted
us to  compute a detailed  grid of evolutionary sequences  for He-core
white dwarfs  appropriate for the  study of these stars  in metal-rich
clusters.  Specifically,  we have computed a total  of 30 evolutionary
sequences  which  cover  a   wide  range  of  stellar  masses.   Three
supersolar metallicities for the progenitor star have been considered:
$Z=0.03$,  0.04, and  0.05.  In  addition, to  study the  evolution of
He-core white dwarfs  with masses larger than the He  core mass at the
He flash  we have also  computed additional sequences  by artificially
suppressing the occurrence of the core He flash at the tip of the RGB.
Details are given  in Table 1.  The evolution of  all white dwarfs has
been followed from the end  of mass-loss episodes during the pre-white
dwarf evolution down to very low surface luminosities.

Our  computations  capture  the  bulk of  the  essential  microphysics
involved in  the evolution of  low-mass white dwarfs  --- particularly
time-dependent  diffusion and  hydrogen  burning ---  as  well as  the
evolutionary  history   of  progenitor  stars   formed  in  metal-rich
environments.  These features allowed us to provide a detailed account
of the  evolutionary properties of  these white dwarfs. The  paper has
also  been intended  to  explore  the role  of  element diffusion  and
residual nuclear burning in the late evolution of He-core white dwarfs
having  supersolar  metallicity.   In  addition  to  the  evolutionary
calculations, we  have computed for all our  sequences accurate colors
and magnitudes  based on new  and improved non-gray  model atmospheres
which explicitly include  Ly$\alpha$ quasi-molecular opacity according
to the  approximation used by Kowalski  \& Saumon (2006).  The grid of
He-core  sequences  presented here  together  with  that presented  in
Serenelli et  al.  (2002) for He-core white  dwarfs of low-metallicity
progenitors  constitute   a  solid   and  consistent  frame   for  the
interpretation  of observations  of  He-core white  dwarfs in  stellar
clusters of different metallicities.

We find  that residual hydrogen  burning strongly influences  the late
stages of white dwarf cooling, substantially delaying the evolution by
several Gyr in some cases.  This  is so despite the fact that the mass
of hydrogen  retained in  the surface layers  at the beginning  of the
cooling stage is markedly smaller  in the case of metal-rich precursor
stars.   The contribution of  residual nuclear  burning to  the energy
budget of  the white  dwarf becomes more  dominat for  smaller stellar
masses.  We arrive  at an opposite conclusion if  element diffusion is
allowed to operate.   Indeed, diffusion processes substantially impact
the evolution of He-core white dwarfs.  The reason for this is that an
important fraction of the hydrogen  content left in the white dwarf is
burnt  during  the  early  stages of  evolution  by  diffusion-induced
hydrogen  burning, with  the consequence  that the  star is  left with
almost  no nuclear energy  at late  stages.  Thus,  diffusion strongly
mitigates the  role of residual nuclear burning,  making the evolution
to rely solely  on the thermal content stored in  the ions, as assumed
in  some  previous  studies  of  low-mass  white  dwarfs  (Althaus  \&
Benvenuto 1997; Hansen \& Phinney 1998).

In view  of the  importance of  diffusion for the  late stages  of the
evolution of He-core  white dwarfs, it is very  important to establish
the occurrence of diffusion in these  stars. It can be argued that the
role of  element diffusion could  by mitigated and even  suppressed by
processes like turbulence or magnetic  fields. In this regard, we find
that a reduction  of the diffusion velocity by a  factor of $\sim 0.7$
is enough to inhibit  the occurrence of the diffusion-induced hydrogen
shell  flashes that  are responsible  for the  fast evolution  at late
stages.  However, there are  numerous and different pieces of evidence
that  sustain the  fact that  diffusion is  an important  process that
occurs both  in the surface layers  and in the deep  interior of white
dwarf stars.   The purity of the  atmosphere of most  white dwarfs and
the observed  spectral evolution are  best understood in terms  of the
occurrence of diffusion.  The existence of some cool white dwarfs with
with helium  atmospheres and traces  of carbon (DQs) are  explained in
terms of the outward diffusion  of the inner carbon profile (Pelletier
et al.  1986). For  low-mass white dwarfs  with helium  cores, element
diffusion  appears to be  also a  viable process  able to  explain the
extremely fast  evolution required to  match the white dwarf  age with
the  spin-down  ages  of   some  millisecond  pulsars  and,  thus,  in
explaining the  observed age dichotomy  (Panei et al 2008;  Althaus et
al. 2001a). In a  different context, asteroseismology can in principle
provide  important information  about the  internal  stratification of
white dwarfs.  For instance, for intermediate-mass white dwarfs it has
been argued (Metcalfe et al.   2005) that diffusion occurs in the deep
envelope  of the  pulsating  He-rich  white dwarf  GD  358.  All  this
evidence hints  at the occurrence  of diffusion processes in  the deep
interior of  white dwarfs and give  strong support to  our result that
residual pp burning  in He-core white dwarfs is not  expected to be an
important  source  of energy  for  significant  periods  of time.   In
closing, we  emphasize that  detailed tabulations of  our evolutionary
sequences  and  colors  for  both  the HST  ACS  filters  and  $UBVRI$
photometric   system   can   be   found   at   our   web   site   {\tt
http://www.fcaglp.unlp.edu.ar/evolgroup}  or   can  be  obtained  upon
request to the authors.

\begin{acknowledgements}
We acknowledge the suggestions and comments of our referee, D. Winget,
which substantially improved the original version of this paper.  This
research  was  supported   by  AGENCIA:  Programa  de  Modernizaci\'on
Tecnol\'ogica   BID  1728/OC-AR,   by  the   AGAUR,  by   MCINN  grant
AYA2008--04211--C02--01, by  the European Union  FEDER funds and  by PIP
6521  grant from  CONICET.  LGA  also acknowledges  AGAUR  through the
Generalitat  de  Catalunya  for   a  PIV  grant.   Finally,  we  thank
R. Martinez and H. Viturro for technical support.
\end{acknowledgements}

\end{document}